\documentclass[aps,prb,reprint,floatfix,superscriptaddress]{revtex4-2}

\usepackage[dvipdfmx]{graphicx}
\usepackage{here}
\usepackage{dcolumn}
\usepackage{bm}
\usepackage{comment}
\usepackage{amsmath}
\usepackage{braket}
\usepackage[colorlinks]{hyperref}

\usepackage{color}

\hypersetup{
	setpagesize=false,
	bookmarksnumbered=true,%
	bookmarksopen=true,%
	colorlinks=true,
	linkcolor=blue,
	citecolor=blue,
	urlcolor=black,
}

\newcommand{\yir}{Y$_{2}$Ir$_{2}$O$_{7}$}
\newcommand{\cdos}{Cd$_{2}$Os$_{2}$O$_{7}$}
\newcommand{\TN}{$T_{\textrm{N}}$}
\newcommand{\aiao}{{all-in/all-out}}
\newcommand{\aoai}{{all-out/all-in}}

\newcommand{\ctext}[1]{\raise0.2ex\hbox{\textcircled{\scriptsize{#1}}}}

\begin{document}
\title{
\texorpdfstring
{Piezomagnetic effect in 5$\bm d$ transition metal oxides \yir\ and \cdos\ with \aiao\ magnetic order}
{Piezomagnetic effect in 5d transition metal oxides Y2Ir2O7 and Cd2Os2O7 with all-in/all-out magnetic order}
}

\author{Hiroki Nanjo}
\email[]{nanjo.hiroki.s7@dc.tohoku.ac.jp}
\affiliation{Department of Physics, Graduate School of Science, Tohoku University, 6-3 Aramaki-Aoba, Aoba-ku, Sendai, Miyagi 980-8578, Japan}
\author{Yoshinori Imai}
\affiliation{Department of Physics, Graduate School of Science, Tohoku University, 6-3 Aramaki-Aoba, Aoba-ku, Sendai, Miyagi 980-8578, Japan}
\author{Takuya Aoyama}
\affiliation{Department of Quantum Matter, Graduate School of Advanced Science and Engineering, Hiroshima University, 1-3-1 Kagamiyama, Higashihiroshima, Hiroshima 735-8530, Japan}
\author{Junichi Yamaura}
\affiliation{The Institute for Solid State Physics, University of Tokyo, 5-1-5 Kashiwanoha, Kashiwa, Chiba 227-8561, Japan}
\author{Kenya Ohgushi}
\affiliation{Department of Physics, Graduate School of Science, Tohoku University, 6-3 Aramaki-Aoba, Aoba-ku, Sendai, Miyagi 980-8578, Japan}

\date{\today}

\begin{abstract}
\section*{Abstract}
{
We investigated the piezomagnetic effect in pyrochlore-type oxides \yir\ and \cdos, which show a non-coplanar magnetic structure called the \aiao \ antiferromagnetic order at low temperatures. The \aiao\ magnetic order can be viewed as a ferroic order of the $xyz$-type magnetic octupoles. We observed a linear development of magnetization with applying stress for both materials. We then estimated the powder-averaged piezomagnetic tensor at 50 K to be $Q = 5.74 \times 10^{-6}$ $\mu_{\text{B}}$/Ir$\cdot$MPa for \yir, and $Q = 3.49 \times 10^{-7}$ $\mu_{\text{B}}$/Os$\cdot$MPa for \cdos. We discuss the microscopic mechanism of the piezomagnetic effect in terms of the stress-induced modification of the g-tensor anisotropy and Dzyaloshinskii-Moriya (D-M) interactions. This work paves the way for the further development of piezomagnetic materials using a strategy based on magnetic multipoles.
}
\end{abstract}

\maketitle

\section*{Introduction}

One of recent advances in the condensed matter physics is the systematic classification of magnetic orders in terms of magnetic multipoles \cite{classificiation_multipole_Hayami, mpg_classification_Yatsushiro}. This methodology is useful since one can systematically predict what kind of cross-correlation responses emerge in a specific magnetic order \cite{cluster_multipole_Suzuki}. For example, the antiferromagnetic states of Co$_4$Nb$_2$O$_9$ \cite{FISCHER_magnetoelectric, Khanh_magnetoelectric, Khanh_electric_polarization, Co4Nb2O9_Yanagi, magnetoelectric_Yanagi} and KOsO$_4$ \cite{magnetoelectric_Hayami, KOsO4_Yamaura} correspond to a ferroic order of magnetic quadrupoles, which generate the magnetoelectric effect. Another example is the antiferromagnetic state of Mn$_3$Sn, in which the anomalous Hall effect and magnetic circular dichroism are argued in relation to the ferroic order of magnetic octupoles \cite{Nakatsuji_AHE, Ikhlas_piezo_AHE}. 

The piezomagnetic effect is a cross-correlation response characterized by a mutual coupling between magnetic and elastic properties. The piezomagnetic effect active materials exhibit the stress-induced magnetization $M_i = Q_{ijk} \sigma_{jk}$ ({$M_i$} being magnetization, and $\sigma_{jk}$ being stress) \cite{classificiation_multipole_Hayami, mpg_classification_Yatsushiro, Ikhlas_piezo_AHE, Dzialoshinskii1958, Moriya1959, Borovik-Romanov1960, Jaime2017, piezo_aoyama}. Here, the piezomagnetic tensor $Q_{ijk}$ is a third-rank axial-$c$ tensor \cite{mpg_classification_Yatsushiro, piezo_aoyama}, and becomes finite when the magnetic dipole, magnetic toroidal quadrupole, and magnetic octupole are ordered in a ferroic manner \cite{classificiation_multipole_Hayami, mpg_classification_Yatsushiro}. In other words, one can explore functional piezomagnetic materials with a knowledge of magnetic multipoles. 

\begin{figure}[t]
  \centering
  \includegraphics[width=\linewidth]{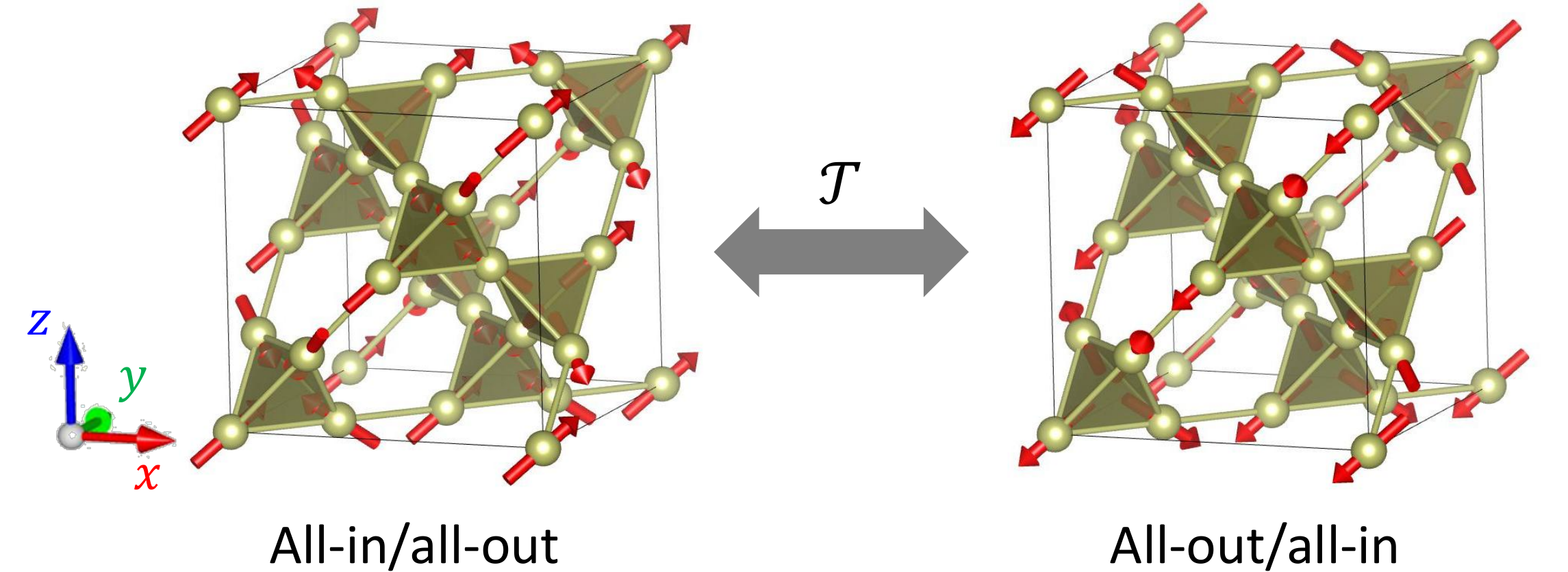}
  \caption{All-in/all-out and  \aoai\ magnetic orders in pyrochlore structure. Both domains are connected by the time reversal operation. This figure is drawn using VESTA \cite{vesta}.}
  \label{Fig1}  
\end{figure}

Pyrochlore-type oxides $A_2B_2$O$_7$ ($A$ = alkaline earth metals or rare earth metals; $B$ = transition metals), which have been extensively studied with a particular interest in geometrical frustration inherent in the tetrahedral network of $A$ and $B$ atoms,  have recently attracted much attention from the viewpoint of higher-order magnetic multipoles \cite{Electric_Toroidal_Quadrupoles_CdOs_Hayami, electric_toroidal_quadrupole_CdOs_Hirose, multipole_227_Nakayama}.  Several pyrochlore oxides such as $A_2$Ir$_2$O$_7$ ($A$ = Y, Nd, Eu) and \cdos\ show the so-called \aiao\ magnetic order, in which  the $A$- and $B$-site spins point towards/opposite to the center of tetrahedron (Fig.\ref{Fig1}) \cite{Ln227_mathuhira_MI, Nd227_tomiyasu, Eu227_sagayama, y227_disseler_exp, y227_direct_evidence, y227_shinaoka, cdos_ohgushi_jpn, cdos_ohgushi}.  The \aiao\ order has the magnetic point group of $m \bar{3}m'$, and can be recognized as the ferroic order of $xyz$-type magnetic octupole. Then, the piezomagnetic tensor of $Q_{xyz}=Q_{yzx}=Q_{zxy}$ becomes finite \cite{aiao_arima, classificiation_multipole_Hayami, Birss1964}. This means that the magnetization along the $z$ direction develops when the shear stress $\sigma_{xy}$ is applied according to the equation $M_z = Q_{zxy} \sigma_{xy}$. The experimental demonstration of the piezomagnetic effect in the \aiao\ magnetic order has been highly expected. 

In this study, we examined the piezomagnetic effect of \yir\ and \cdos, which shows an \aiao\ magnetic order at low temperatures. \yir, in which Ir$^{4+}$ ions have  $(5d)^5$ electron configuration, is a spin-orbital Mott insulator characterized by the localized $j_{\text{eff}} = 1/2$ state \cite{Wan_yir_configure, Clancy_yir_configure}.  The material undergoes an antiferromagnetic transition at \TN\ = 151 K. In contrast, \cdos, in which Os$^{5+}$ ions have  $(5d)^3$ electron configuration,  is located close to the metal-insulator transition \cite{Sala_CaIrO3_configure, Matsuda_cdos_XMCD,  Bogdanov_cdos_distortion}. This compound undergoes a metal-insulator transition that is driven by antiferromagnetic ordering at \TN\ = 227 K. We observed the linear development of magnetization with respect to the applied stress, and revealed that the averaged piezomagnetic tensor at 50 K are $Q = 5.74 \times 10^{-6}$ $\mu_{\text{B}}$/Ir$\cdot$MPa for \yir, and $ Q = 3.49 \times 10^{-7}$ $\mu_{\text{B}}$/Os$\cdot$MPa for \cdos. We discuss the microscopic mechanism of the stress-induced magnetization in terms of the modification of g-tensor anisotropy and Dzyaloshinskii-Moriya (D-M) interactions due to the symmetry reduction of the crystal structure.

\section*{Experiment}
Polycrystalline samples of \yir\ were synthesized by a standard solid-state reaction method.  A mixture of Y$_2$O$_3$ and IrO$_2$ in their stoichiometric ratios were heated in air and in a sealed quartz tube at 1373 K for a total of 238 hours with several intermediate grindings.  Polycrystalline samples of \cdos\ were synthesized from CdO and Os with a ratio of 1.05:1 in a sealed quartz tube under the supply of an appropriate amount of AgO as the oxygen source at 1073 K \cite{cdos_Hiroi, Hirose_Nature}. 

The piezomagnetic effect was evaluated by measuring the magnetization under the applied stress. The powder samples are put into a piston cylinder made of Duratron and the stress is applied by the hydraulic pump. The piston was clamped with a non-magnetic Cu-Be cell to maintain stress \cite{piezo_aoyama}. The magnetization of the cell is about $10^{-5}$ emu, which is sufficiently small compared to the magnetization of the samples. The magnetizations are measured by using the commercial magnetometer (MPMS, Quantum Design) under selected magnetic fields. A typical measurement was done by applying magnetic field of $H_{\text{meas}} = $ 10 Oe.

In order to observe the piezomagnetic effect, one has to get a single-domain state, since the distinct antiferromagnetic order connected by the time-reversal operation has an opposite sign of the $Q_{ijk}$ tensor \cite{classificiation_multipole_Hayami}.  In our case, \aiao\ and \aoai\ orders correspond to two antiferromagnetic domains (Fig. \ref{Fig1}).  A single-domain state is realized by cooling the sample across \TN\ under applying stress and magnetic field. This is enabled by the difference in the free energy $F = Q_{ijk} H_i \sigma_{jk}$ between two antiferromagnetic domains. A typical poling field is $H_{\text{pol}}$ = 10000 Oe in our experiments. The magnetization was measured during heating process after this poling process. 

Another important issue is that our experiments were performed for polycrystalline samples. In this case, each crystal is randomly oriented, forming an effectively isotropic distribution. When uniaxial stress is applied to polycrystals, one can measure the powder-averaged piezomagnetic signals. The powder-averaged of the piezomagnetic tensor can be calculated by  
\begin{equation*}
Q^{\text{ave}}_{ijk} = \frac{\int_{0}^{2 \pi} d \phi \int_{0}^{\pi} d \theta \sin{\theta} \int_{0}^{2 \pi} d \psi |A_{ii'}^{-1} A_{jj'}^{-1} A_{kk'}^{-1} Q_{i'j'k'}|}{8 \pi^2}.
\end{equation*}
Here, $A_{ij}$ represents the rotation matrix, which transforms the local crystallographic coordinates to the laboratory coordinates.  Our experimental setup detects the $Q^{\text{ave}}_{zzz}$ component.  The calculation for the \aiao\ order leads to $Q^{\text{ave}}_{zzz} = \frac{3 }{4 \pi}Q_{xyz}$. We hereafter call this quantity as $Q$ for simplicity.

\section*{Results}

\begin{figure}[t]
  \centering
  \includegraphics[width=\linewidth]{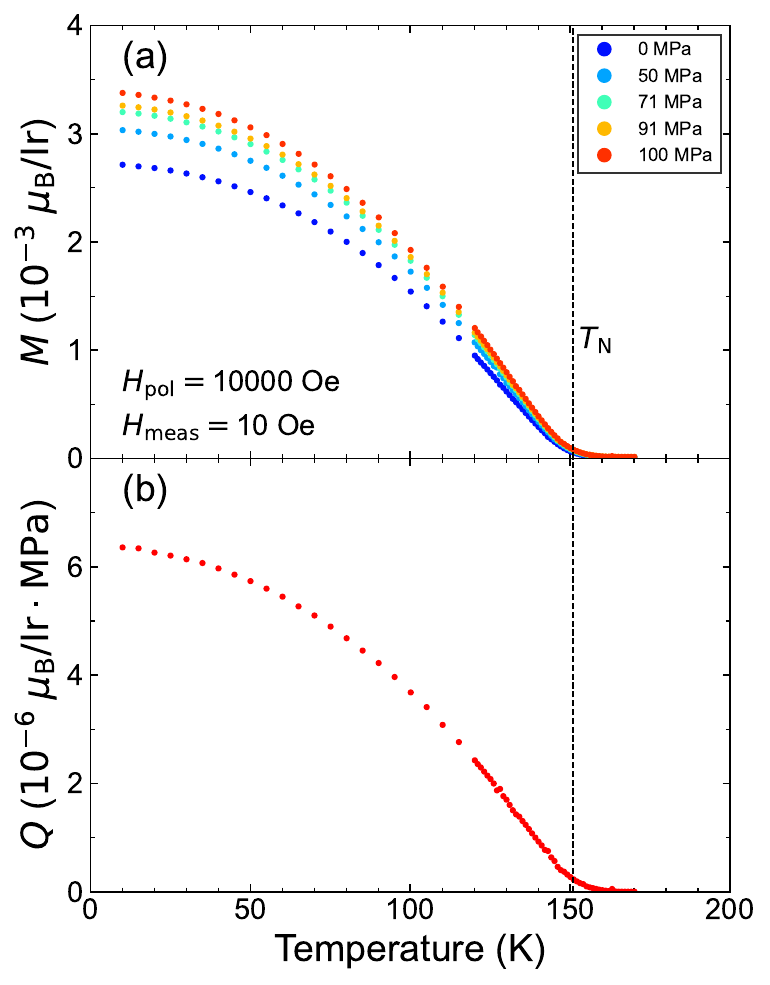}
  \caption{(a) Temperature dependence of magnetization for \yir\ under various stresses. Measurements were performed at $H_{\text{meas}}=10$ Oe after cooling the sample at $H_{\text{pol}}=10000$ Oe. The background data coming from the cell assembly has been subtracted from the data in advance. (b) Temperature dependence of the powder-averaged piezomagnetic tensor $Q$ for \yir.}
  \label{Fig2}
\end{figure}

\begin{figure}[t]
  \centering
  \includegraphics[width=\linewidth]{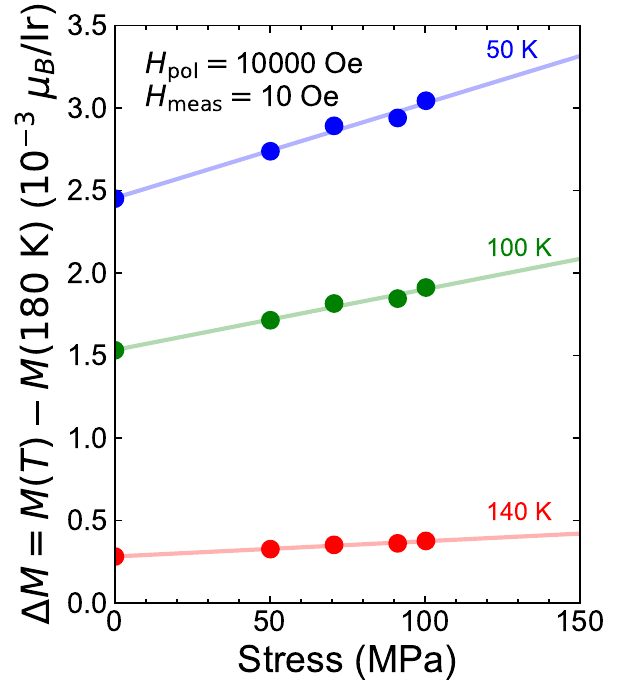}
  \caption{The stress dependence of $\Delta M = M(T)-M$(180 K), which is the relative change in the magnetization from the paramagnetic phase for \yir\ at 50, 100 and 140 K. The solid line indicates the results of fitting with $\Delta M=M_0 +Q \sigma$, where $M_0$ stands for the spontaneous magnetization and $\sigma$ is the applied stress.}
  \label{Fig3}
\end{figure}

Figure \ref{Fig2}(a) shows the temperature ($T$) dependence of magnetization ($M$) for \yir\ under various stresses ($\sigma$). Poling procedure was performed at $H_{\text{pol}} = $ 10000 Oe, and the magnetization was measured at $H_{\text{meas}} = $ 10 Oe on a heating process. At $\sigma  =0$ MPa, the spontaneous magnetization develops below the antiferromagnetic transition temperature \TN\ = 151 K. It reaches 2.71 $\times 10^{-3}$ $\mu_{\text{B}}$/Ir at the lowest temperature measured. On applying stress, the spontaneous magnetization below \TN\ increases monotonically, indicating a close coupling between magnetic and elastic properties. The observed magnetization subtracted by the value at 180 K in the paramagnetic state, $\Delta M = M(T)- M$(180 K), is plotted against the stress in Fig. \ref{Fig3}. One can see that magnetization increases linearly with applying stress, which is the characteristic of the piezomagnetic effect. From fitting the data with an equation $\Delta M =M_0 + Q \sigma$, we obtain the spontaneous magnetization $M_0 = 2.45 \times 10^{-3}$ $\mu_{\text{B}}$/Ir and the powder-averaged piezomagnetic tensor $Q = 5.74 \times 10^{-6}$ $\mu_{\text{B}}$/Ir$\cdot$MPa for \yir\ at 50 K. We repeated the same fitting at each temperature and get the temperature dependence of $Q$, which is presented in Fig. \ref{Fig2}(b). One can see that $Q$ increases monotonically with decreasing temperature. 

\begin {figure}[t]
  \centering
  \includegraphics[width=\linewidth]{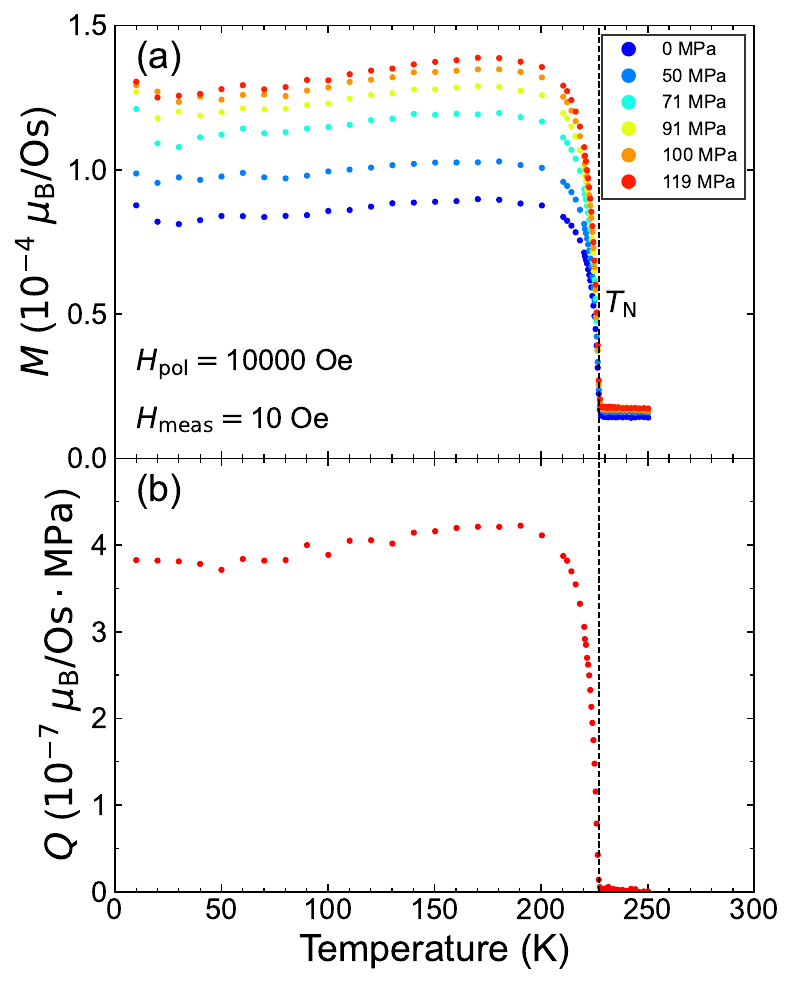}
  \caption{(a) Temperature dependence of magnetization for \cdos\ under various stresses. Measurements were performed at $H_{\text{meas}}=10$ Oe after cooling the sample at $H_{\text{pol}}=10000$ Oe. The background data coming from the cell assembly has been subtracted from the data in advance. (b) Temperature dependence of the powder-averaged piezomagnetic tensor $Q$ for \cdos.}
  \label{Fig4}
\end{figure}

\begin{figure}[t]
  \centering
  \includegraphics[width=\linewidth]{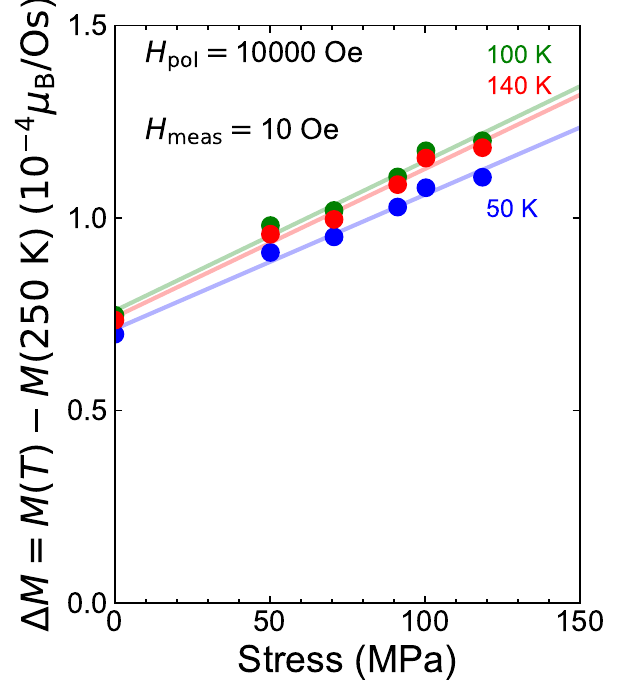}
  \caption{The stress dependence of $\Delta M = M(T)-M$(250 K), which is the relative change in the magnetization from the paramagnetic phase for \cdos\ at 50, 100 and 140 K. The solid line indicates the results of fitting with $\Delta M=M_0 +Q \sigma$, where $M_0$ stands for the spontaneous magnetization and $\sigma$ is the applied stress.}
  \label{Fig5}
\end{figure}

Figure \ref{Fig4}(a) shows the temperature dependence of magnetization for \cdos\ under various stresses. Poling process was performed at $H_{\text{pol}} = $ 10000 Oe, and the magnetization was measured at $H_{\text{meas}} = $ 10 Oe on heating process. At $\sigma  =0$ MPa, the spontaneous magnetization develops below the antiferromagnetic transition temperature \TN\ = 227 K. It reaches 8.77 $\times 10^{-5}$ $\mu_{\text{B}}$/Os at the lowest temperature measured. On applying stress, the spontaneous magnetization below \TN\ increases monotonically. The observed magnetization subtracted by the value at 250 K in the paramagnetic state, $\Delta M = M(T)- M$(250 K), is plotted against the stress in Fig. \ref{Fig5}. One can see that magnetization increases linearly with applying stress, which is the characteristic of the piezomagnetic effect. From fitting the data with an equation $\Delta M =M_0 + Q \sigma$, we obtain the spontaneous magnetization $M_0 = 7.11 \times 10^{-5}$ $\mu_{\text{B}}$/Os and the powder-averaged piezomagnetic tensor $Q = 3.49 \times 10^{-7}$ $\mu_{\text{B}}$/Os$\cdot$MPa for \cdos\ at 50 K. We repeated the same fitting at each temperature and get the temperature dependence of $Q$, which is presented in Fig. \ref{Fig4}(b). One can see that the $Q$ values sharply increase at \TN\ and keeps almost constant below \TN. This behavior is different from the behavior of $Q$ for \yir.

Summarizing experimental results, we successfully observed the stress-induced magnetization in both \yir\ and \cdos. The linear development of magnetization with respect to applied stress is the direct evidence of the piezomagnetic effect generated by the magnetic octupoles. The magnitude of the powder-averaged piezomagnetic tensor for \yir\ is about 10 times larger than that for \cdos.  This difference is likely originating from the distinct electron configurations between Ir$^{4+}$ and Os$^{5+}$ ions. We also note that the piezomagnetic tensor of $Q = 5.74 \times 10^{-6}$ $\mu_{\text{B}}$/Ir$\cdot$MPa for \yir\ is rather large among known piezomagnetic materials; the value is approximately half of the $Q$ value of CoF$_2$, which is known to have a large piezomagnetic effect \cite{Borovik-Romanov1960}. The reason is likely relevant to a large spin-orbit coupling among $5d$ systems. Another important observation is the weak ferromagnetism observed in both \yir\ and \cdos. We note that the irreducible representation containing \aiao\ order does not contain any ferromagnetic components \cite{cdos_ohgushi_jpn, cdos_ohgushi}; hence, emergence of weak ferromagnetism should not be allowed from the symmetrical point of view.  In the following sections, we discuss the microscopic mechanism of the piezomagnetic effect and weak ferromagnetism.

\section*{Discussion}
\subsection{The microscopic mechanism of the piezomagnetic effect}

\begin{figure*}[t]
  \centering
  \includegraphics[width=\linewidth]{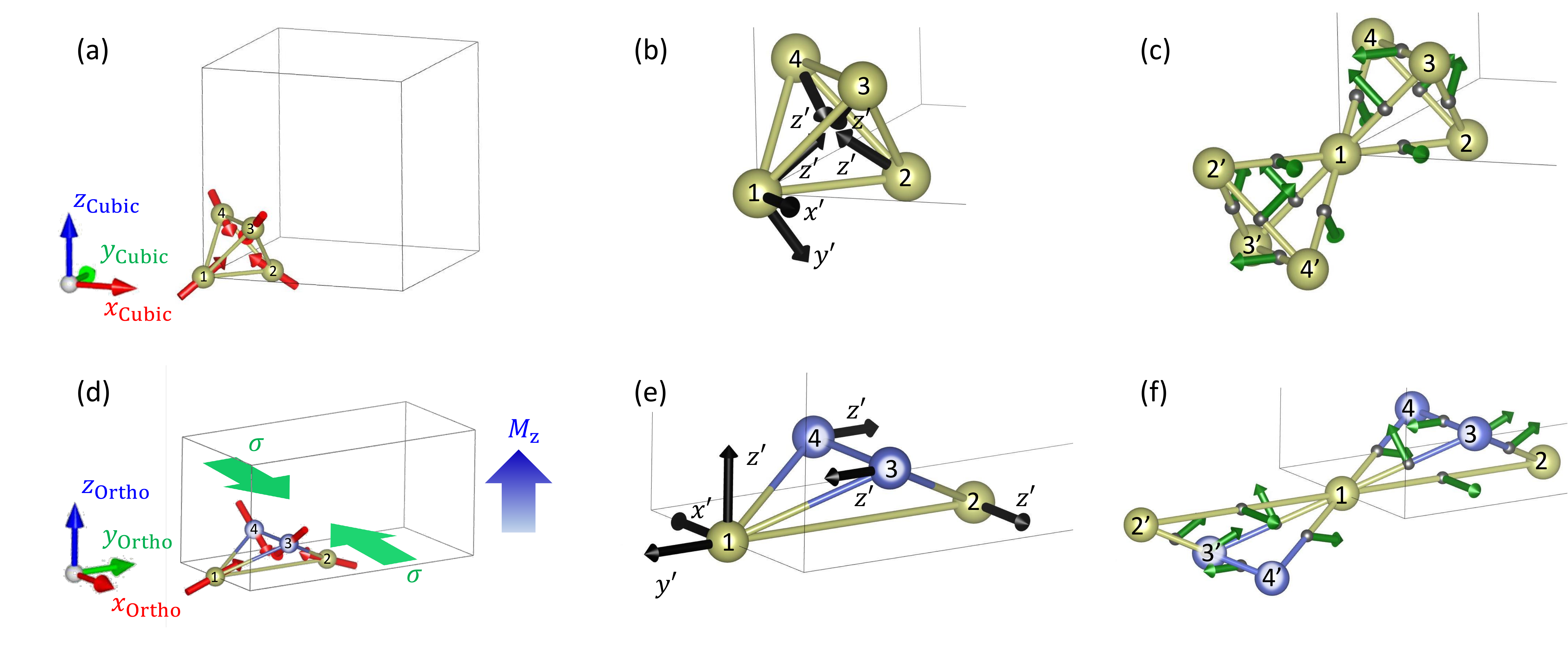}
  \caption{(a) All-in/all-out magnetic order in the pyrochlore structure with the cubic $Fd\bar{3}m$ symmetry. The spin directions are $\bm{S_1} = \frac{S}{\sqrt{3}}(1, 1, 1)$ at $\bm{r_1} = (0, 0, 0)$, $\bm{S_2} = \frac{S}{\sqrt{3}}\left(-1, -1, 1\right)$ at $\bm{r_2} = \left(\frac{1}{4}, \frac{1}{4}, 0\right)$, $\bm{S_3} = \frac{S}{\sqrt{3}}\left(-1, 1, -1\right)$ at $\bm{r_3} = \left(\frac{1}{4}, 0, \frac{1}{4}\right)$, and $\bm{S_4} = \frac{S}{\sqrt{3}}\left(1, -1, -1\right)$ at $\bm{r_4} = \left(0, \frac{1}{4}, \frac{1}{4}\right)$ in the cubic coordinate. (b) The local principal axes ($z'$ axes), which are indicated by black arrows, at four $B$-sites in the $Fd\bar{3}m$ symmetry. (c) The D-M vectors, which are indicated by green arrows, at twelve midpoints in the cubic $Fd\bar{3}m$ symmetry. (d) All-in/all-out magnetic order in the pyrochlore structure with the orthorhombic $Imma$ symmetry. The spin directions are $\bm{S_1} = \frac{S}{\sqrt{3}}\left(0, \sqrt{2}, 1\right)$ at $\bm{r_1} = \left(\frac{1}{2}, 0, 0\right)$, $\bm{S_2} = \frac{S}{\sqrt{3}}\left(0, -\sqrt{2}, 1\right)$ at $\bm{r_2} = \left(\frac{1}{2}, \frac{1}{2}, 0\right)$, $\bm{S_3} = \frac{S}{\sqrt{3}}\left(-\sqrt{2}, 0, -1\right)$ at $\bm{r_3} = \left(\frac{3}{4}, \frac{1}{4}, \frac{1}{4}\right)$, and $\bm{S_4} = \frac{S}{\sqrt{3}}\left(\sqrt{2}, 0, -1\right)$ at $\bm{r_4} = \left(\frac{1}{4}, \frac{1}{4}, \frac{1}{4}\right)$. The green arrows indicate the applied stress $\sigma_{xy}$.  The blue arrow expresses the magnetization induced by the piezomagnetic effect $M_z = Q_{zxy} \sigma_{xy}$. (e) The local principal axes ($z'$ axes), which are indicated by black arrows, at four $B$-sites in the orthorhombic $Imma$ symmetry. (f) The D-M vectors, which are indicated by green arrows, at twelve midpoints in the orthorhombic $Imma$ symmetry.}
  \label{Fig_discussion}
\end{figure*}

In the \aiao\ magnetic structure with the $m\bar{3}m'$ magnetic point group, there are three piezomagnetic tensor components $Q_{xyz}, Q_{yzx}$, and $Q_{zxy}$. Among them, the $Q_{zxy}$ component indicates that the magnetization along the $z$ direction is generated by applying the shear stress $\sigma_{xy}$, which is written as $M_z=Q_{zxy}\sigma_{xy}$ (Fig. \ref{Fig_discussion}(a), (d)). In this section, we discuss the microscopic mechanism of the piezomagnetic effect in this case. The $Q_{xyz}$ and $Q_{yzx}$ components can be understood in a similar manner; actually, there is a relationship of $Q_{xyz}=Q_{yzx}=Q_{zxy}$. The piezomagnetic effect is interpreted in two steps: (1) the applied stress induces the lattice distortions and lowers the crystal symmetry, and  (2)  the magnitude and/or direction of magnetic moments changes accordingly, generating the macroscopic magnetization \cite{Moriya1959}. We here describe the first step. 

Applying the $\sigma_{xy}$ stress is equivalent to the compression in the [1-10] direction, which induces the change in the space group from the cubic $Fd\bar{3}m$ (Fig. \ref{Fig_discussion}(a)) to the orthorhombic $Imma$ symmetry (Fig. \ref{Fig_discussion}(d)). Then, four $B$-atoms located at the $16c$ sites (labelled by 1, 2, 3 and 4 in Fig. \ref{Fig_discussion}(a)) in the cubic $Fd\bar{3}m$ symmetry are split into two $B$-atoms located at the $4b$ sites (labelled by 1 and 2 in Fig. \ref{Fig_discussion}(d)) and two $B$-atoms located at the $4c$ sites (labelled by 3 and 4 in Fig. \ref{Fig_discussion}(d)) in the orthorhombic $Imma$ symmetry. This symmetry reduction results in the emergence of uniform magnetic moments in two mechanisms: (i) g-tensor anisotropy and (ii) D-M interactions. These two mechanisms are detailed in the following subsections.

\subsubsection{g-tensor anisotropy mechanism}

The g-tensor is restricted by the local environment of the magnetic ions. Let us consider the g-tensor in the cubic $Fd\bar{3}m$ symmetry. The four $B$-atoms occupy the $16c$ sites with the site symmetry of $\bar{3}m$. Then, the g-tensor with the rank-2 polar-$i$ nature at $n$-th site can be expressed as  
\begin{equation*}
\bm{g_n^{\mathrm{local}}}=
\begin{pmatrix}
g_{xx} & 0 & 0 \\
  0 & g_{xx} & 0 \\
  0 & 0 & g_{zz} \\
\end{pmatrix}
\ (n = 1 \sim 4), 
\end{equation*}
where the local principal axis ($z'$ axis) is along a unit vector of   
$\bm{n_1^{\mathrm{cubic}}}=\frac{1}{\sqrt{3}}(1, 1, 1)$, 
$\bm{n_2^{\mathrm{cubic}}}=\frac{1}{\sqrt{3}}(-1, -1, 1)$, 
$\bm{n_3^{\mathrm{cubic}}}=\frac{1}{\sqrt{3}}(-1, 1, -1)$, and  
$\bm{n_4^{\mathrm{cubic}}}=\frac{1}{\sqrt{3}}(1, -1, -1)$ (Fig. \ref{Fig_discussion}(b)) \cite{Birss1964}. We note that the cubic coordinate shown in Fig. \ref{Fig_discussion}(a) is adopted here. The total magnetization per unit cell is expressed as $\bm{M} = \sum_{n=1}^4 \mu_{\text{B}} \bm{g_n^{\mathrm{global}}} \cdot \bm{S_n}$, where the g-tensor in the global coordinate $\bm{g_n^{\mathrm{global}}} = \bm{A^{-1}} \cdot \bm{g_n^{\mathrm{local}}} \cdot \bm{A}$ with a suitable rotation matrix $\bm{A}$ is introduced. 
Putting the \aiao\ magnetic order with 
$\bm{S_1} = \frac{S}{\sqrt{3}} \left(1, 1, 1\right), 
\bm{S_2} = \frac{S}{\sqrt{3}} \left(-1, -1, 1\right), 
\bm{S_3} = \frac{S}{\sqrt{3}} \left(-1, 1, -1\right), 
\bm{S_4} = \frac{S}{\sqrt{3}} \left(1, -1, -1\right)$ into the above formula, 
we obtain $\bm{M} = (0, 0, 0)$ for the cubic crystal. 

When the pyrochlore structure is distorted into the orthorhombic $Imma$ symmetry, $B$-atoms are split into two sets: two atoms located at the $4b$ sites with  $2/m$ site symmetry and two atoms located at the $4c$ sites with the $2/m$ site symmetry. The g-tensor at the $4b$ sites can be expressed as  
\begin{equation*}
\begin{split}
\bm{g_n^{\mathrm{local}}}=
\begin{pmatrix}
g_{xx}^a & g_{xy}^a & 0 \\
g_{yx}^a & g_{yy}^a & 0 \\
  0 & 0 & g_{zz}^a \\
\end{pmatrix}
\ (n = 1, 2),  \\
\end{split}
\end{equation*}
where the local principal axis ($z'$ axis) is along a unit vector of   
$\bm{n_1^{\mathrm{ortho}}}=(-1, 0, 0)$, and 
$\bm{n_2^{\mathrm{ortho}}}=(1, 0, 0)$ (Fig. \ref{Fig_discussion}(e)) . We note that the orthorhombic coordinate shown in Fig. \ref{Fig_discussion}(d) is adopted here. 
Similarly, the g-tensor at the $4c$ sites can be expressed as
\begin{equation*}
\begin{split}
\bm{g_n^{\mathrm{local}}}=
\begin{pmatrix}
g_{xx}^b & g_{xy}^b & 0 \\
g_{yx}^b & g_{yy}^b & 0 \\
  0 & 0 & g_{zz}^b \\
\end{pmatrix}
\ (n = 3, 4), \\
\end{split}
\end{equation*}
where the local principal axis ($z'$ axis) is along a unit vector of $\bm{n_3^{\mathrm{ortho}}}=(0, -1, 0)$, and $\bm{n_4^{\mathrm{ortho}}}=(0, 1, 0)$ (Fig. \ref{Fig_discussion}(e)) . We note that the orthorhombic coordinate shown in Fig. \ref{Fig_discussion}(d) is adopted here. The total magnetization per unit cell in the \aiao\ magnetic order is calculated to be 
\begin{equation*}
\begin{split}
\bm{M} &= \sum_{n = 1}^4 \mu_{\text{B}} \bm{g_n^{\mathrm{global}}} \cdot \bm{S_n} \\
&= \frac{2(-\sqrt{6}g_{yx}^a - \sqrt{6}g_{yx}^b + \sqrt{3}g_{yy}^a - \sqrt{3}g_{yy}^b)}{3} \mu_{\text{B}} S
\begin{pmatrix}
0 \\
0 \\
1\\
\end{pmatrix}
\end{split}
\end{equation*}
for the orthorhombic crystal. The magnetization develops in the $z$-axis when the shear stress $\sigma_{xy}$ is applied. We note that the induced magnetization becomes zero, when the cubic conditions are recovered. The cubic conditions are $g_{xx}^a = \left( g_{xx} + 2 g_{zz} \right)/3, \ g_{xy}^a = \sqrt{2}\left(g_{xx} - g_{zz}\right)/3, \ g_{yx}^a = \sqrt{2}\left(g_{xx} - g_{zz}\right)/3, \ g_{yy}^a = \left(2 g_{xx} + g_{zz}\right)/3, \ g_{zz}^a = g_{xx}, \ g_{xx}^b = \left(g_{xx} + 2 g_{zz}\right)/3, \ g_{xy}^b = -\sqrt{2}\left(g_{xx} - g_{zz}\right)/3, \ g_{yx}^b = -\sqrt{2}\left(g_{xx} - g_{zz}\right)/3, \ g_{yy}^b = \left(2 g_{xx} + g_{zz}\right)/3$, and $g_{zz}^b = g_{xx}$. These are totally consistent with the symmetry considerations, and this g-tensor scenario well explains the piezomagnetic effect in \aiao\ magnetic structure in pyrochlore oxides.

\subsubsection{D-M interactions mechanism}
We next focus on the symmetry at the midpoint of magnetic ions, which influences the D-M interactions. In the cubic $Fd\bar{3}m$ symmetry, the midpoint of two nearest-neighbor magnetic ions is located at $48f$ sites with the $2mm$ site symmetry. Therefore, there are two mirror planes: one is perpendicular to the bond, and the other is perpendicular to the first plane and contains the bond. According to Moriya’s rule, D-M vector between the two magnetic ions must be perpendicular to the mirror plane that contains the bond. All the D-M vectors in a unit cell be written using a single parameter $D$ as
$\bm{D_{12}}=D(1, -1, 0)/\sqrt{2}$, 
$\bm{D_{13}}=D(-1, 0, 1)/\sqrt{2}$, 
$\bm{D_{14}}=D(0, 1, -1)/\sqrt{2}$, 
$\bm{D_{23}}=D(0, -1, -1)/\sqrt{2}$, 
$\bm{D_{24}}=D(1, 0, 1)/\sqrt{2}$, 
$\bm{D_{34}}=D(-1, -1, 0)/\sqrt{2}$
$\bm{D_{12'}}=D(1, -1, 0)/\sqrt{2}$, 
$\bm{D_{13'}}=D(-1, 0, 1)/\sqrt{2}$, 
$\bm{D_{14'}}=D(0, 1, -1)/\sqrt{2}$, 
$\bm{D_{2'3'}}=D(0, -1, -1)/\sqrt{2}$, 
$\bm{D_{2'4'}}=D(1, 0, 1)/\sqrt{2}$, and
$\bm{D_{3'4'}}=D(-1, -1, 0)/\sqrt{2}$
 \cite{Krempa_DM}, where we use the cubic coordinate (Fig. \ref{Fig_discussion}(c)).

We now consider the Hamiltonian 
\begin{equation}\label{eq:Hamiltonian}
\mathcal{H} = \sum_{\langle i, j \rangle}J  \bm{S_i} \cdot \bm{S_j} + \sum_{\langle i, j \rangle} \bm{D_{ij}} \cdot ( \bm{S_i} \times \bm{S_j}),
\end{equation}
where $J(>0)$ is the Heisenberg-type exchange interactions, and is much larger than the D-M interactions, $J \gg D$. 
We here suppose that the spin state can be written as 
\begin{equation*}
  \begin{split}
  \bm{S_1} = \frac{S}{\sqrt{3}}
  \begin{pmatrix}
  \cos{\theta} - \frac{1}{\sqrt{2}} \sin{\theta} \\
  \cos{\theta} - \frac{1}{\sqrt{2}} \sin{\theta} \\
  \sqrt{2} \sin{\theta} + \cos{\theta} 
  \end{pmatrix} \\
  \bm{S_2} = \bm{S_{2'}} = \frac{S}{\sqrt{3}}
  \begin{pmatrix}
  -\cos{\theta} + \frac{1}{\sqrt{2}} \sin{\theta}  \\
  -\cos{\theta} + \frac{1}{\sqrt{2}} \sin{\theta}\\
  \sqrt{2} \sin{\theta} + \cos{\theta} 
  \end{pmatrix}\\
  \bm{S_3} = \bm{S_{3'}} = \frac{S}{\sqrt{3}}
  \begin{pmatrix}
  -\cos{\theta} - \frac{1}{\sqrt{2}} \sin{\theta}  \\
  \cos{\theta} + \frac{1}{\sqrt{2}} \sin{\theta}  \\
  \sqrt{2} \sin{\theta} - \cos{\theta} 
  \end{pmatrix} \\
  \bm{S_4} = \bm{S_{4'}} = \frac{S}{\sqrt{3}}
  \begin{pmatrix}
  \cos{\theta} + \frac{1}{\sqrt{2}} \sin{\theta}  \\
  -\cos{\theta} - \frac{1}{\sqrt{2}} \sin{\theta}  \\
  \sqrt{2} \sin{\theta} - \cos{\theta} 
  \end{pmatrix},
  \end{split}
  \end{equation*}
which expresses the spin canting from the \aiao\ magnetic order. The spin directions are determined so that the energy obtained from the Hamiltonian $\mathcal{H} $ is minimized. We then see that the energy becomes minimum when $\theta =0$, which is nothing but the \aiao\ order. 

When the pyrochlore structure is distorted into orthorhombic $Imma$ symmetry, the twelve midpoints in a unit cell are split into three sets: four midpoints located at the two $4e$ sites with $mm2$ site symmetry and eight midpoints located at the $16j$ sites with the $1$  site symmetry. 
In the orthorhombic crystal, four mirror planes are retained at midpoints 1-2, 3-4, 1-$2'$, and $3'$-$4'$ with $mm2$ site symmetry,  while mirror planes are broken at the other midpoints 1-3, 1-4, 2-3, 2-4, 1-$3'$, 1-$4'$, $2'$-$3'$, and $2'$-$4'$ with $1$ site symmetry. Then, D-M vectors are determined to be 
$\bm{D_{12}}=D'(1, 0, 0)$, 
$\bm{D_{13}}=(D_x, D_y, D_z)$, 
$\bm{D_{14}}=(D_x, -D_y, -D_z)$, 
$\bm{D_{23}}=(-D_x, D_y, -D_z)$, 
$\bm{D_{24}}=(-D_x, -D_y, D_z)$, 
$\bm{D_{34}}=D''(0, -1, 0)$
$\bm{D_{12'}}=D'(1, 0, 0)$, 
$\bm{D_{13'}}=(D_x, D_y, D_z)$, 
$\bm{D_{14'}}=(D_x, -D_y, -D_z)$, 
$\bm{D_{2'3'}}=(-D_x, D_y, -D_z)$, 
$\bm{D_{2'4'}}=(-D_x, -D_y, D_z)$, and
$\bm{D_{3'4'}}=D''(0, -1, 0)$, where we use the orthorhombic coordinate (Fig. \ref{Fig_discussion}(f)). 

We now consider the Hamiltonian Eq.\eqref{eq:Hamiltonian}, where we suppose that the Heisenberg-type exchange interactions are the same as in cubic $Fd\bar{3}m$ symmetry. We also suppose that the spin state can be written as
\begin{equation*}\label{eq:spin_cant_ortho}
  \begin{split}
  \bm{S_1} = \frac{S}{\sqrt{3}}
  \begin{pmatrix}
  0 \\
  -\sin{\theta} + \sqrt{2} \cos{\theta} \\
  \sqrt{2} \sin{\theta} + \cos{\theta} 
  \end{pmatrix} \\
  \bm{S_2} = \bm{S_{2'}} = \frac{S}{\sqrt{3}}
  \begin{pmatrix}
  0 \\
  \sin{\theta} - \sqrt{2} \cos{\theta} \\
  \sqrt{2} \sin{\theta} + \cos{\theta} 
  \end{pmatrix}\\
  \bm{S_3} = \bm{S_{3'}} = \frac{S}{\sqrt{3}}
  \begin{pmatrix}
  -\sin{\theta} - \sqrt{2} \cos{\theta} \\
  0 \\
  \sqrt{2} \sin{\theta} - \cos{\theta} 
  \end{pmatrix} \\
  \bm{S_4} = \bm{S_{4'}} = \frac{S}{\sqrt{3}}
  \begin{pmatrix}
  \sin{\theta} + \sqrt{2} \cos{\theta} \\
  0 \\
  \sqrt{2} \sin{\theta} - \cos{\theta} 
  \end{pmatrix},
  \end{split}
  \end{equation*}
which expresses the spin canting from the \aiao\ magnetic order in the orthorhombic coordinate. Then, we get the total energy of 
\begin{equation*}
\begin{split}
E = &\frac{32JS^2}{3} \sin^2{\theta} + \frac{2}{3} S^2 (D' - D'' + 6 D_x - 6 D_y) \cos{2\theta} \\
 & +\frac{4S^2}{3} \left( \sqrt{2} D' + \sqrt{2} D'' -3 D_z \right) \sin{2 \theta} + const.
\end{split}
\end{equation*}
The total energy is minimized when 
\begin{equation*}
\theta = \theta_c \sim \frac{-D' + D'' - 6 D_x + 6D_y}{16J},
\end{equation*}
from which we obtain the net magnetization as 
\begin{equation*}
\bm{M} = \sum_{n=1}^{4} g \mu_{\text{B}} \bm{S_n} \sim \frac{\sqrt{6}}{12} \frac{-D' + D'' - 6 D_x + 6D_y}{J}
\begin{pmatrix}
0 \\
0 \\
1 \\
\end{pmatrix}
\end{equation*}
for a orthorhombic crystal.  These results indicate that the magnetization develops in the $z$-axis when the shear stress $\sigma_{xy}$ is applied to the pyrochlore structure. We note that the induced magnetization becomes zero, when the cubic conditions are recovered. The cubic conditions are $D' = D'' = D, \  D_x = -D/\sqrt{2}, \ \  D_y = -D/\sqrt{2}$, and $\  D_z = D/\sqrt{2}$. These are totally consistent with the symmetry considerations, and this D-M interaction scenario well explains the piezomagnetic effect in \aiao\ magnetic structure in pyrochlore oxides.

\subsection{The microscopic mechanism of the weak ferromagnetism}
Finally, we discuss the microscopic mechanism of weak ferromagnetism observed in both \yir\ and \cdos. We recall that spontaneous magnetization is prohibited in the irreducible representation containing the \aiao\ order. The earlier study argues that weak ferromagnetism emerges inside the magnetic domain walls. The magnetic domain wall between the \aiao\ and \aoai\ magnetic order consists of tetrahedra with 2-in/2-out magnetic order, in which an uncompensated magnetic moment remains perpendicular to the magnetic domain wall \cite{Hirose_Nature}. This scenario is supported by experiments by Hiroi \textit{et al}., who showed that the weak ferromagnetic moment scales with the polycrystalline grain size \cite{cdos_Hiroi}. The coexistence of two magnetic domains has been visualized in \cdos\ by the circular polarized resonant X-ray diffraction imaging technique \cite{Tardif_cdos_imaging}. 

We here propose another scenario with particular attention on the piezomagnetic effect. We suppose that there are tiny but sizable crystal distortions, which possibly introduced by the Jahn-Teller instability inherent in unfilled nature of $5d$ orbitals of Ir$^{4+}$ and Os$^{5+}$ ions. When the crystal structure is spontaneously deformed, magnetization becomes finite even in the absence of stress  owing to “the spontaneous piezomagnetic effect”. Based on the measured values of the piezomagnetic tensor $Q \sim 10^{-6}$ $\mu_{\text{B}}$/Ir$\cdot$MPa and weak ferromagnetic moment $M_{0} \sim 10^{-3}$ $\mu_{\text{B}}$/Ir, one can estimate the self-induced stress to be $\sigma \sim 10^{3}$ MPa. Putting the elastic modulus of $C \sim 300$ GPa, which is a typical value for the pyrochlore-type oxides \cite{Pruneda_elastic_constant, Qu_elastic_constant}, into the relationship $\sigma = C \epsilon$, one can expect the self-induced shear strain of $\epsilon \sim 10^{-3}$. The observation of this tiny shear strain by means of high-resolution X-ray diffraction is highly expected to verify this scenario.

\section*{CONCLUSION}
In this study, we observed the piezomagnetic effect, which is the stress-induced magnetization, in \yir\ and \cdos\ with \aiao\ magnetic order. The powder-averaged piezomagnetic tensor at 50 K is $5.74 \times 10^{-6}$ $\mu_{\text{B}}$/Ir$\cdot$MPa for \yir\ and $3.49 \times 10^{-6}$ $\mu_{\text{B}}$/Os$\cdot$MPa for \cdos, indicating that the magnitude of piezomagnetic tensor of \yir\ is about 10 times larger than that of \cdos. The observed piezomagnetic effect originates from a ferroic order of magnetic octupoles from the macroscopic viewpoint. Microscopically, the piezomagnetic effect is found to originate from the modification of the g-tensor anisotropy and the D-M interactions, which is induced by the symmetry reduction in the crystal structure. We also observed weak ferromagnetism in antiferromagnetic states, which can be explained by the magnetic domain wall and/or the spontaneous lattice distortion scenario. Our findings demonstrate the potential power of a multipole strategy for designing functional piezomagnetic materials.

\bibliography{Ref_lists.bib}

\begin{thebibliography}{10}

\bibitem{classificiation_multipole_Hayami}
S.~Hayami, M.~Yatsushiro, Y.~Yanagi, and H.~Kusunose, Phys. Rev. B {\bfseries 98},  165110 (2018).

\bibitem{mpg_classification_Yatsushiro}
M.~Yatsushiro, H.~Kusunose, and S.~Hayami, Phys. Rev. B {\bfseries 104},  054412 (2021).

\bibitem{cluster_multipole_Suzuki}
M.-T. Suzuki, T.~Koretsune, M.~Ochi, and R.~Arita, Phys. Rev. B {\bfseries 95},  094406 (2017).

\bibitem{FISCHER_magnetoelectric}
E.~Fischer, G.~Gorodetsky, and R.~Hornreich, Solid State Communications {\bfseries 10},  1127 (1972).

\bibitem{Khanh_magnetoelectric}
N.~D. Khanh, N.~Abe, H.~Sagayama, A.~Nakao, T.~Hanashima, R.~Kiyanagi, Y.~Tokunaga, and T.~Arima, Phys. Rev. B {\bfseries 93},  075117 (2016).

\bibitem{Khanh_electric_polarization}
N.~D. Khanh, N.~Abe, S.~Kimura, Y.~Tokunaga, and T.~Arima, Phys. Rev. B {\bfseries 96},  094434 (2017).

\bibitem{Co4Nb2O9_Yanagi}
Y.~Yanagi, S.~Hayami, and H.~Kusunose, Physica B: Condensed Matter {\bfseries 536},  107 (2018).

\bibitem{magnetoelectric_Yanagi}
Y.~Yanagi, S.~Hayami, and H.~Kusunose, Phys. Rev. B {\bfseries 97},  020404 (2018).

\bibitem{magnetoelectric_Hayami}
S.~Hayami, H.~Kusunose, and Y.~Motome, Phys. Rev. B {\bfseries 97},  024414 (2018).

\bibitem{KOsO4_Yamaura}
J.~Yamaura and Z.~Hiroi, Phys. Rev. B {\bfseries 99},  155113 (2019).

\bibitem{Nakatsuji_AHE}
S.~Nakatsuji, N.~Kiyohara, and T.~Higo, Nature {\bfseries 527},  212 (2015).

\bibitem{Ikhlas_piezo_AHE}
M.~Ikhlas, S.~Dasgupta, F.~Theuss, T.~Higo, S.~Kittaka, B.~J. Ramshaw, O.~Tchernyshyov, C.~W. Hicks, and S.~Nakatsuji, Nature Physics {\bfseries 18},  1086 (2022).

\bibitem{Dzialoshinskii1958}
I.~Dzialoshinskii, Sov. Phys. JETP {\bfseries 6},  621 (1958).

\bibitem{Moriya1959}
T.~Moriya, J. Phys. Chem. Solids {\bfseries 11},  73 (1959).

\bibitem{Borovik-Romanov1960}
A.~S. Borovik-Romanov, J. Exptl. Theoret. Phys. (U.S.S.R.) {\bfseries 11},  1088 (1960).

\bibitem{Jaime2017}
M.~Jaime, A.~Saul, M.~Salamon, V.~S. Zapf, N.~Harrison, T.~Durakiewicz, J.~C. Lashley, D.~A. Andersson, C.~R. Stanek, J.~L. Smith, and K.~Gofryk, Nat. Commun. {\bfseries 8},  1 (2017).

\bibitem{piezo_aoyama}
T.~Aoyama and K.~Ohgushi, Phys. Rev. Mater. {\bfseries 8},  L041402 (2024).

\bibitem{Electric_Toroidal_Quadrupoles_CdOs_Hayami}
S.~Hayami, Y.~Yanagi, H.~Kusunose, and Y.~Motome, Phys. Rev. Lett. {\bfseries 122},  147602 (2019).

\bibitem{electric_toroidal_quadrupole_CdOs_Hirose}
H.~T. Hirose, T.~Terashima, D.~Hirai, Y.~Matsubayashi, N.~Kikugawa, D.~Graf, K.~Sugii, S.~Sugiura, Z.~Hiroi, and S.~Uji, Phys. Rev. B {\bfseries 105},  035116 (2022).

\bibitem{multipole_227_Nakayama}
Y.~Nakayama, D.~Hirai, H.~Sagayama, K.~Kojima, N.~Katayama, J.~Lehmann, Z.~Wang, N.~Ogawa, and K.~Takenaka, Phys. Rev. Mater. {\bfseries 8},  055001 (2024).

\bibitem{Ln227_mathuhira_MI}
K.~Matsuhira, M.~Wakeshima, Y.~Hinatsu, and S.~Takagi, Journal of the Physical Society of Japan {\bfseries 80},  094701 (2011).

\bibitem{Nd227_tomiyasu}
K.~Tomiyasu, K.~Matsuhira, K.~Iwasa, M.~Watahiki, S.~Takagi, M.~Wakeshima, Y.~Hinatsu, M.~Yokoyama, K.~Ohoyama, and K.~Yamada, Journal of the Physical Society of Japan {\bfseries 81},  034709 (2012).

\bibitem{Eu227_sagayama}
H.~Sagayama, D.~Uematsu, T.~Arima, K.~Sugimoto, J.~J. Ishikawa, E.~O'Farrell, and S.~Nakatsuji, Phys. Rev. B {\bfseries 87},  100403 (2013).

\bibitem{y227_disseler_exp}
S.~M. Disseler, C.~Dhital, A.~Amato, S.~R. Giblin, C.~de~la Cruz, S.~D. Wilson, and M.~J. Graf, Phys. Rev. B {\bfseries 86},  014428 (2012).

\bibitem{y227_direct_evidence}
S.~M. Disseler, Phys. Rev. B {\bfseries 89},  140413 (2014).

\bibitem{y227_shinaoka}
H.~Shinaoka, S.~Hoshino, M.~Troyer, and P.~Werner, Phys. Rev. Lett. {\bfseries 115},  156401 (2015).

\bibitem{cdos_ohgushi_jpn}
J.~Yamaura, K.~Ohgushi, and Z.~Hiroi, Nihon Kessho Gakkaishi {\bfseries 55},  116 (2013).

\bibitem{cdos_ohgushi}
J.~Yamaura, K.~Ohgushi, H.~Ohsumi, T.~Hasegawa, I.~Yamauchi, K.~Sugimoto, S.~Takeshita, A.~Tokuda, M.~Takata, M.~Udagawa, M.~Takigawa, H.~Harima, T.~Arima, and Z.~Hiroi, Phys. Rev. Lett. {\bfseries 108},  247205 (2012).

\bibitem{aiao_arima}
T.~Arima, Journal of the Physical Society of Japan {\bfseries 82},  013705 (2013).

\bibitem{Birss1964}
R.~R. Birss.
\newblock \textit{Symmetry and Magnetism}, (North-Holland Publishing Company, 1964).

\bibitem{Wan_yir_configure}
X.~Wan, A.~M. Turner, A.~Vishwanath, and S.~Y. Savrasov, Phys. Rev. B {\bfseries 83},  205101 (2011).

\bibitem{Clancy_yir_configure}
J.~P. Clancy, N.~Chen, C.~Y. Kim, W.~F. Chen, K.~W. Plumb, B.~C. Jeon, T.~W. Noh, and Y.-J. Kim, Phys. Rev. B {\bfseries 86},  195131 (2012).

\bibitem{Sala_CaIrO3_configure}
M.~Sala, K.~Ohgushi, A.~Al-Zein, Y.~Hirata, G.~Monaco, and M.~Krisch, Physical Review Letters {\bfseries 112},  (2014).

\bibitem{Matsuda_cdos_XMCD}
Y.~H. Matsuda, J.~L. Her, S.~Michimura, T.~Inami, M.~Suzuki, N.~Kawamura, M.~Mizumaki, K.~Kindo, J.~Yamauara, and Z.~Hiroi, Phys. Rev. B {\bfseries 84},  174431 (2011).

\bibitem{Bogdanov_cdos_distortion}
N.~A. Bogdanov, R.~Maurice, I.~Rousochatzakis, J.~van~den Brink, and L.~Hozoi, Phys. Rev. Lett. {\bfseries 110},  127206 (2013).

\bibitem{vesta}
K.~Momma and F.~Izumi, J. Appl. Cryst. {\bfseries 44},  1272 (2011).

\bibitem{cdos_Hiroi}
Z.~Hiroi, J.~Yamaura, T.~Hirose, I.~Nagashima, and Y.~Okamoto, APL Materials {\bfseries 3},  041501 (2015).

\bibitem{Hirose_Nature}
H.~T. Hirose, J.~Yamaura, and Z.~Hiroi, Scientific Reports {\bfseries 7},  42440 (2017).

\bibitem{Krempa_DM}
W.~Witczak-Krempa and Y.~B. Kim, Phys. Rev. B {\bfseries 85},  045124 (2012).

\bibitem{Tardif_cdos_imaging}
S.~Tardif, S.~Takeshita, H.~Ohsumi, J.~Yamaura, D.~Okuyama, Z.~Hiroi, M.~Takata, and T.~Arima, Phys. Rev. Lett. {\bfseries 114},  147205 (2015).

\bibitem{Pruneda_elastic_constant}
J.~M. Pruneda and E.~Artacho, Phys. Rev. B {\bfseries 72},  085107 (2005).

\bibitem{Qu_elastic_constant}
Z.~Qu, C.~Wan, and W.~Pan, Acta Materialia {\bfseries 60},  2939 (2012).

\end{thebibliography}
\bibliographystyle{prb.bst}

\end{document}